\documentclass[sigconf,anonymous=False,review=False]{acmart}
\AtBeginDocument{%
  }

\usepackage{hhline,colortbl}
\usepackage[dvipsnames]{xcolor}
\usepackage{subcaption}

\setcopyright{acmlicensed}
\copyrightyear{2026}
\acmYear{2026}
\acmDOI{XXXXXXX.XXXXXXX}
\acmConference[Conference acronym 'XX]{Make sure to enter the correct
  conference title from your rights confirmation email}{June 03--05,
  2018}{Woodstock, NY}
\acmISBN{978-1-4503-XXXX-X/2018/06}

\begin{document}

\title[Towards Gaze-Informed AI Disclosure Interfaces]{Towards Gaze-Informed AI Disclosure Interfaces: Eye-Tracking Attentional and Cognitive Load While Reading AI-Assisted News}

\author{Pooja Prajod}
\email{Pooja.Prajod@cwi.nl}
\orcid{0000-0002-3168-3508}
\affiliation{%
  \institution{Centrum Wiskunde \& Informatica}
  \city{Amsterdam}
  \country{The Netherlands}
}

\author{Hannes Cools}
\affiliation{%
  \institution{University of Amsterdam}
  \city{Amsterdam}
  \country{The Netherlands}
}

\author{Thomas R\"oggla}
\affiliation{%
  \institution{Centrum Wiskunde \& Informatica}
  \city{Amsterdam}
  \country{The Netherlands}
}

\author{Pablo Cesar}
\affiliation{%
  \institution{Centrum Wiskunde \& Informatica}
  \city{Amsterdam}
  \country{The Netherlands}
  }
  \additionalaffiliation{%
  \institution{TU Delft}
  \city{Delft}
  \country{The Netherlands}
  }

\author{Abdallah El Ali}
\affiliation{%
  \institution{Centrum Wiskunde \& Informatica}
  \city{Amsterdam}
  \country{The Netherlands}
  }
  \additionalaffiliation{%
  \institution{Utrecht University}
  \city{Utrecht}
  \country{The Netherlands}
  }

\renewcommand{\shortauthors}{Prajod et al.}

\begin{abstract}

As generative AI becomes increasingly integrated into journalism, designing effective AI-use disclosures that inform readers without imposing unnecessary burden is a key challenge. While prior research has primarily focused on trust and credibility, the impact of disclosures on readers' attentional and cognitive load remains underexplored. To address this gap, we conducted a $3\times2\times2$ mixed factorial study manipulating the level of AI-use disclosure detail (none, one-line, detailed), news type (politics, lifestyle), and role of AI (editing, partial content generation), measuring load via NASA-TLX and eye-tracking. Our results reveal a significant attentional cost: one-line disclosures resulted in significantly higher fixation durations and saccade counts, particularly for AI-edited content. Detailed disclosures did not impose additional burden. Drawing on Information-Gap Theory, we argue that brief labels may trigger increased visual scrutiny by alerting readers to AI use without providing enough information. NASA-TLX scores and pupil diameter showed no significant differences across conditions, suggesting that AI-use disclosures do not impose cognitive burden regardless of the detail level. Interview insights contextualize these findings and reveal a strong preference for detailed or ``detail-on-demand'' designs. Our findings inform the design of gaze-informed adaptive disclosure interfaces that dynamically adjust transparency levels based on readers' attentional patterns and news context.

\end{abstract}



\begin{teaserfigure}
\centering
  \includegraphics[width=0.95\textwidth]{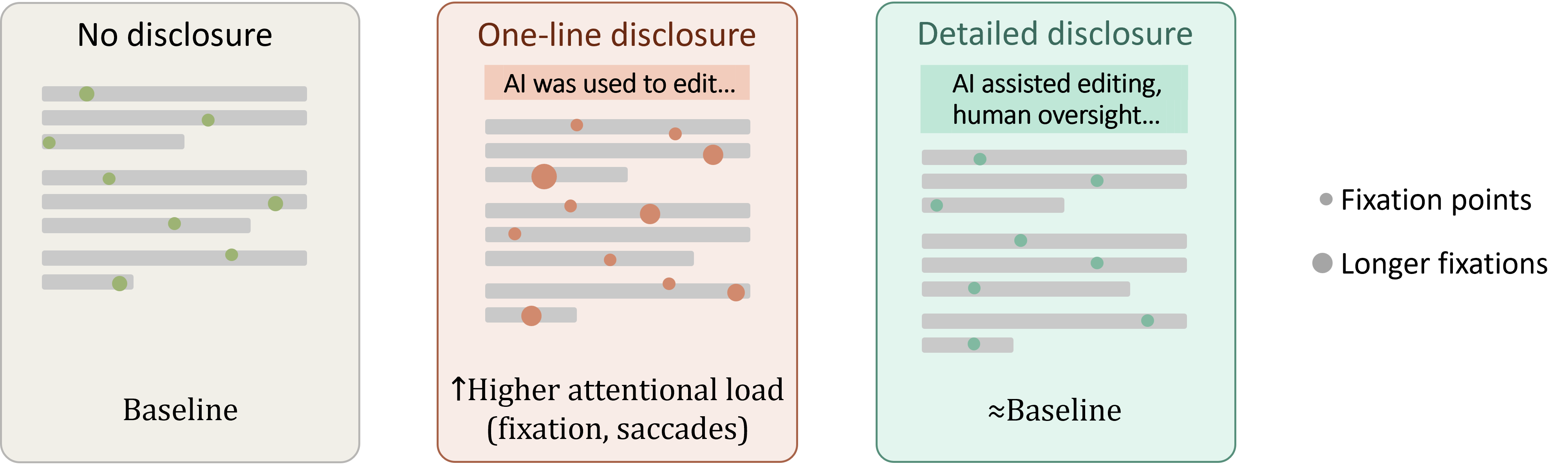}
  \caption{Illustration of key findings: one-line AI-use disclosures increase readers' attentional load (larger, more frequent fixation points) compared to no disclosure and detailed disclosure conditions.}
  \Description{Three simulated news article cards representing the three disclosure conditions. The no-disclosure card shows sparse, small fixation dots indicating baseline attentional load. The one-line disclosure card shows larger and more numerous fixation dots, indicating higher attentional load. The detailed disclosure card shows sparse fixation dots similar to baseline, despite containing more disclosure text.}
  \label{fig:teaser}
\end{teaserfigure}

\iffalse
\received{20 February 2007}
\received[revised]{12 March 2009}
\received[accepted]{5 June 2009}
\fi

\maketitle

\section{Introduction}

As generative AI becomes increasingly integrated into news production, there are growing calls for transparency regarding its use~\cite{zier2024labeling, el2024transparent}. While disclosure statements are the primary mechanism for communicating AI involvement to readers, a key open challenge~\cite{springer2020progressive} is \textit{designing disclosures that inform readers about AI use without imposing attentional and cognitive load.} Recent advancements in physiological computing suggest that real-time gaze sensing can enable interfaces to detect and adapt to such use states~\cite{kosch2023survey, prajod2023gaze, chiossi2024physiochi}, which is a promising path towards user-centric, dynamic news interfaces. Given this potential, we investigate whether gaze signals can reveal when a disclosure design increases attentional and cognitive load during news reading.

Current AI-use disclosures in news are largely self-regulated and non-standardized~\cite{becker2025policies, zier2024labeling}, ranging from minimal one-line labels to detailed transparency statements~\cite{altay2024people, toff2025or, prajod2026full, morosoli2025transparency}. AI's role in content production also varies, from editorial assistance to content generation~\cite{kusters2026human, gilardi2024willingness, valenzuela2026effects}. To date, research has focused almost exclusively on trust and credibility~\cite{altay2024people, nanz2025ai, morosoli2025transparency, toff2025or, longoni2022news, prajod2026full, leuppert2025ai}, and the resulting reader engagement, such as sharing and subscription. However, the attentional and cognitive costs of AI disclosures remain underexplored. This gap is significant for news interface design as higher cognitive load can reduce information retention and increase susceptibility to misinformation~\cite{apuke2024information, debue2014does}. Understanding these effects requires real-time sensing methods, as subjective measures alone may not capture implicit processing differences~\cite{chiossi2024physiochi, liu2026sensing}.

We bridge this gap by exploring two research questions central to the design of adaptive news interfaces: 
\begin{itemize}
    \item \textbf{Does the level of detail in AI disclosures influence readers' attentional and cognitive load?}
    \item \textbf{Does attentional and cognitive load differ depending on AI's role (editing vs. partial content generation)?}
\end{itemize}
 
We address these through a 3$\times$2$\times$2 mixed factorial study by manipulating AI disclosure detail (none, one-line, detailed), news type (politics, lifestyle), and AI role (editing, partial content generation). We measured load via NASA-TLX~\cite{hart1988development} and eye-tracking (pupil diameter, fixation duration, saccade count). NASA-TLX and pupil diameter capture cognitive load~\cite{kosch2023survey}, while fixation duration and saccades reflect attentional load (visual search and spatial attention distribution)~\cite{skaramagkas2021review}.

Our results reveal an attentional load cost for one-line disclosures, i.e., one-line disclosures increase attentional load, particularly for AI-edited content, whereas detailed disclosures do not impose additional burden despite containing more information. Drawing on Information-Gap Theory~\cite{loewenstein1994psychology}, we argue that brief labels may alert readers to AI use without providing enough information to resolve the reader's uncertainty, thereby triggering increased visual scrutiny as a form of information-seeking behavior. On the other hand, NASA-TLX scores and pupil diameter showed no significant differences across conditions, indicating that cognitive load remains unchanged regardless of disclosure detail. 

Crucially, eye-tracking captured the attentional load differences, which highlights the value of real-time gaze sensing for designing and evaluating disclosure designs. Our findings inform the design of gaze-informed adaptive disclosure interfaces that dynamically adjust the detail level based on readers' attentional patterns and news context. 

\section{Related Work}
\subsection{Eye-Tracking Measures of Attentional and Cognitive Load}
\label{sec:background}
Eye-tracking measures are increasingly popular for assessing cognitive and attentional load in HCI~\cite{kosch2023survey, ke2024using, barz2024humaneyeze}. In particular, pupil diameter is governed by the autonomic nervous system and is a well-established indicator of cognitive load, with larger pupil sizes reflecting higher cognitive effort~\cite{skaramagkas2021review}. 
Fixations, or focused gaze on areas of interest, and saccades, which are rapid eye movements between fixations, can be voluntary or involuntary. They are also associated with cognitive processing, but are more directly reflective of how attention is distributed across an interface.
Longer fixation durations indicate greater processing difficulty at the point of gaze~\cite{skaramagkas2021review, de2023reading}, while higher saccade counts reflect increased visual search and re-reading behavior~\cite{skaramagkas2021review}. Although subjective measures such as NASA-TLX capture perceived task load, non-intrusive measures like eye-tracking are particularly valuable as they capture real-time, implicit responses that subjective measures may miss~\cite{heimerl2024fordigitstress, liu2026sensing, prajod2024flow}. Moreover, real-time gaze measures have been proposed for adaptive interfaces~\cite{plopski2022eye, menges2019improving, sun2026eyes, barz2024humaneyeze, alhargan2017multimodal, lavit2024gaze, murakami2025inferring, barral2020eye} and hence, eye-tracking is a crucial modality to consider for adaptive AI disclosures.

\subsection{Attentional and Cognitive Load in News Reading}

Recent works have used eye-tracking to investigate cognitive load in news reading contexts. In the misinformation domain, readers show longer fixation durations and increased pupil diameter when reading fake news compared to real news~\cite{hansen2020factuality, sumer2021fakenewsperception, shi2023true, gupta2020eyes}. For instance, an eye-tracking study~\cite{shi2023true} demonstrated the varying cognitive load and gaze patterns when reading true and false COVID-19 news headlines.

Recent works have also studied attentional and cognitive load in the context of AI-generated news content using eye-tracking and physiological measures~\cite{ilyas2025reading, wu2025characterizing, zhang2024you}. These studies generally report higher load when reading AI-generated content compared to human-written content. In an eye-tracking study~\cite{ilyas2025reading} comparing AI-generated and human-written texts, longer fixation durations were observed for AI-generated content, suggesting greater processing difficulty. This observation was not explained by differences in readability scores, leading the authors to attribute it to subtle stylistic differences. A multimodal study combining gaze and physiological measures explored how readers engage with human- and AI-generated real and fake news articles~\cite{wu2025characterizing}. While readers' accuracy at distinguishing AI-generated from human-written articles was near chance level, machine learning models trained on physiological and gaze data were able to predict content source, indicating that implicit processing differences exist even when readers cannot explicitly identify them.  Differences in psychophysiological responses to emotional AI-written news have been observed among young readers~\cite{zhang2024you}.

Notably, the above works focus on the load induced by fake news or AI-generated content itself, rather than by AI-use disclosures.

\begin{table*}
  \caption{AI disclosures statements used in the study.}
  \label{tab:disclosures}
\begin{center}
%\vspace{-10pt}
    \begin{tabular}{p{17.4cm}}
    \toprule
    %\cellcolor{gray!15} \\[-10pt]
    \multicolumn{1}{c}{\cellcolor{gray!15}
    \textbf{Disclosures for AI in editing role}}\\
    \textbf{\underline{One-line:}} An AI tool was used to add an extra layer to the editing process of this article just before it was published.\\
    \\[-7pt]
    \textbf{\underline{Detailed:}} This article was produced with the assistance of an AI tool, which was used to support various stages of the editorial process, including content structuring, language refinement, and fact-check suggestions. All content was reviewed, edited, and approved by a human journalist before publication. This use of AI aligns with our commitment to transparency and responsible innovation in journalism. You can report errors at: webhomepage@news.com. \\ 
    \midrule
    \multicolumn{1}{c}{\cellcolor{gray!15}
    \textbf{Disclosures for AI in partial content generation role}} \\
    \textbf{\underline{One-line:}} This article was largely generated with the help of an AI tool.\\
    \\[-7pt]
    \textbf{\underline{Detailed:}} This article was primarily generated using an AI tool, which was responsible for drafting substantial portions of the content, including initial story development, writing, and factual synthesis. Human editors reviewed the final version prior to publication. Our use of AI is guided by principles of transparency, accountability, and editorial oversight. Readers can report errors at: webhomepage@news.com. \\ 
    \bottomrule
    \end{tabular}

\end{center}  
\end{table*}

\subsection{AI-Use Disclosures in News Reading}
AI-use disclosures in news are currently self-regulated and non-standardized~\cite{becker2025policies, zier2024labeling}. Regulatory frameworks such as the EU AI Act propose transparency obligations for AI systems, but editorial processes and content generation typically fall outside their scope, leaving journalistic AI use largely unaddressed by explicit regulation~\cite{helberger2023european, zier2024labeling}. In practice, disclosure formats vary widely across news organizations, ranging from brief one-line labels to detailed transparency statements describing the specific production steps in which AI was involved~\cite{prajod2026full, NordicAI, gamage2025labeling}. Crucially, effective disclosure design requires balancing informativeness with usability~\cite{el2024transparent, kusters2026human}.

Prior research on AI-use disclosures in news has focused primarily on trust and credibility outcomes. Several studies have observed a ``transparency dilemma'' where disclosing AI involvement paradoxically reduces readers' trust~\cite{morosoli2025transparency, altay2024people, nanz2025ai, toff2025or}. A recent study~\cite{altay2024people} found that labeling headlines as AI-generated reduced perceived accuracy regardless of the headline's veracity. The authors attributed this to AI-aversion caused by readers assuming fully AI-generated content. Similarly, studies reported that AI-generated news is believed less than human-written news~\cite{longoni2022news}. Prior work has also observed differential effects based on news type, with political news being more sensitive to disclosure effects than lifestyle or entertainment news~\cite{morosoli2025transparency, nanz2025ai}. A study~\cite{nanz2025ai} highlighted that AI-generated news outlets were trusted less, particularly for political news, with potential economic implications for subscription willingness. On the other hand, AI-role disclosures can be used to improve readers' short-term engagement with the article~\cite{gilardi2024willingness}.

Recent work has begun exploring how disclosure design choices affect these outcomes. Studies investigating label designs for AI-generated content on social media demonstrated that more detailed labels can improve transparency perceptions~\cite{gamage2025labeling, chen2025examining}. In contrast, detailed labels in a news context can lead to reduced trust, including on measures like willingness to subscribe~\cite{prajod2026full}. Interestingly, different visual disclosure designs can lead to varying perceptions of AI contributions, even when the detail level is the same~\cite{kusters2026human}.

While these studies provide valuable insights into how disclosures affect trust and credibility, the attentional and cognitive load implications of AI-use disclosures remain largely unexamined. Understanding these implications is critical for designing disclosures that inform readers without imposing unnecessary processing burden, particularly given evidence that higher cognitive load can reduce information retention and increase misinformation sharing~\cite{apuke2024information, debue2014does}. One EEG study~\cite{liu2023nudging} found that AI-generated content labels increased attention and cognitive processing compared to unlabeled content. While this suggests that AI disclosure labels affect neural processing, the study used a single binary label (AI-generated vs. no label) and did not examine how the level of detail in disclosures or the role of AI in content production shape readers' attentional and cognitive load. We address this gap through an eye-tracking study that varies disclosure detail and AI role. Moreover, our findings contribute empirical evidence toward gaze-informed disclosure design that balances transparency with readers' cognitive and attentional load.

\section{Methods}

Prior work on AI disclosures has observed differential effects on trust and engagement based on the level of disclosure detail~\cite{hossain2025effects, ngo2025balancing, prajod2026full, springer2020progressive}, the type of news~\cite{nanz2025ai, morosoli2025transparency, prajod2026full, gamage2025labeling}, and the role of AI in content production~\cite{gamage2025labeling, gilardi2024willingness, kusters2026human}. This motivates investigating whether these factors similarly influence readers' attentional and cognitive load. We therefore conducted a lab study using a $3 \times 2 \times 2$ mixed factorial design to explore the impact of AI-use disclosures on readers' attentional and cognitive load. The study manipulated AI disclosure detail (between-subjects: no disclosure, one-line disclosure, detailed disclosure), news type (within-subjects: political, lifestyle), and AI role (within-subjects: editing, partial content generation).

\subsection{News Articles and AI-use Disclosures}

%The news stimuli were based on six actual news articles (three from politics and three from lifestyle) from mainstream news organizations (e.g., BBC, CNN, NOS). Each article was manipulated using ChatGPT-4o to create two AI-assisted versions, resulting in a total of 12 stimuli articles. The AI-edited versions were created using the prompt: \texttt{\textbf{\color{black!70}{Generate one short news article around 250-300 words based solely on the article. Make very small tweaks in the main text based on `SOURCE'. You are allowed to change the title.}}} The resulting articles were minimally edited to ensure consistency with the source material. The partially AI-generated versions were created using the prompt: \texttt{\textbf{\color{black!70}{Generate one short news article around 250-300 words, have a title, an introduction and specific text on `TOPIC'. Make sure it is in the form of coherent article that is based on the following URL: `SOURCE'.}}} The generated articles were used without additional edits, except for deleting a paragraph when necessary to maintain the target word count.

The news stimuli were based on six actual news articles (three from politics and three from lifestyle) from mainstream news organizations (e.g., BBC, CNN, NOS). Each article was manipulated using ChatGPT-4o to create two AI-assisted versions, resulting in 12 stimuli articles. AI-edited versions were generated by prompting the model to make minimal tweaks to the source article while preserving its structure: \texttt{\textbf{\color{black!70}{Generate one short news article around 250-300 words based solely on the article. Make very small tweaks in the main text based on `SOURCE'. You are allowed to change the title.}}} The resulting articles were minimally edited to ensure consistency with the source material. Partially AI-generated versions were created by prompting the model to generate a new 250-300 word article based on the source article: \texttt{\textbf{\color{black!70}{Generate one short news article around 250-300 words, have a title, an introduction and specific text on `TOPIC'. Make sure it is in the form of coherent article that is based on the following URL: `SOURCE'.}}} The generated articles were used without additional human edits, except for deleting a paragraph when necessary to maintain the target word count.

We implemented two versions of AI disclosures: one-line and detailed, both worded to reflect AI's role (see Table~\ref{tab:disclosures}). These statements followed existing AI transparency guidelines for journalism~\cite{BBC, NordicAI, valenzuela2026effects}, and were phrased to avoid implying fully AI-generated content, as prior work indicates this can cause AI aversion~\cite{altay2024people}. Following insights from previous works~\cite{kusters2026human, longdisclosure}, disclosures were displayed at the top of the article as illustrated in Figure~\ref{fig:teaser}. The wording and presentation of the disclosures were pilot-tested with a small group (N=4) of participants. 

\begin{table}
  \caption{Readability scores per article.}
  \label{tab:readability}
  \begin{tabular}{l|cc|cc}
    \toprule
     & \multicolumn{2}{c|}{AI-edited} & \multicolumn{2}{c}{Partially AI-generated} \\
     \hline
     & Flesch & ARI & Flesch & ARI \\
    \hline
    Political 1 & 27.2 & 17.87 & 23.2 & 17.22 \\
    Political 2 & 42.2 & 14.64 & 18.5 & 19.82 \\
    Political 3 & 28.3 & 14.22 & 10.0 & 17.79 \\
    Lifestyle 1 & 45.7 & 12.67 & 38.7 & 12.99 \\
    Lifestyle 2 & 47.3 & 13.26 & 29.8 & 16.11 \\
    Lifestyle 3 & 50.4 & 13.01 & 38.4 & 13.68 \\
    \hline
    \textbf{Mean} & \textbf{40.18} & \textbf{14.28} & \textbf{26.43} & \textbf{16.27} \\
    \bottomrule
  \end{tabular}
\end{table}

To characterize the linguistic complexity of the stimuli, we computed two readability metrics for each article (Table~\ref{tab:readability}). The Flesch Reading Ease score~\cite{flesch1948new} estimates readability based on sentence length and syllable count, with higher scores indicating easier text. The Automated Readability Index (ARI)~\cite{senter1967automated} estimates the grade level required to comprehend the text based on character and word counts, with lower scores indicating easier text. As shown in Table~\ref{tab:readability}, AI-edited articles were consistently easier to read than partially AI-generated articles across both metrics. Lifestyle articles were also easier to read than political articles.

\begin{figure*}
    \centering
    \includegraphics[width=0.85\linewidth]{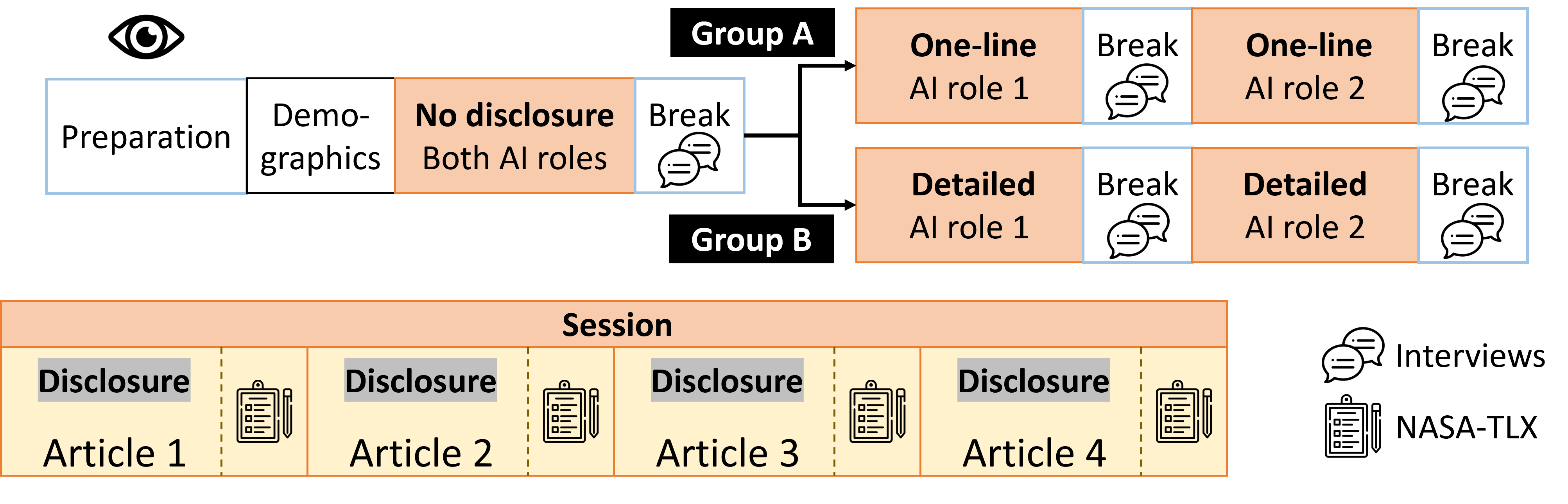}
    %\vspace{-10pt}
    \caption{Overview of the experimental protocol. The eye tracker was calibrated during preparation and recorded gaze data throughout the experiment. AI role 1 and AI role 2 refer to editing and partial generation, counterbalanced across participants.}
    \label{fig:loadprotocol}
    \Description[Protocol for group A and B]{The initial part of the study is the same for both groups. Session 2 and 3 for group A is with one-line disclosure, and for group B is with detailed disclosure.}
\end{figure*}

\subsection{Measures}

We used both subjective and objective measures to capture attentional and cognitive load in response to AI-use disclosures.

\textbf{NASA-TLX:} We used an adapted version of the NASA-TLX scale~\cite{hart1988development} for subjective assessment of perceived task load. NASA-TLX measures perceived load across six sub-scales, but we focused on mental demand, effort, and frustration, as our study did not involve physical exertion, time pressure, or competitive elements. Participants rated each sub-scale on a 10-point scale after reading each article, resulting in a combined score out of 30.

\textbf{Eye-Tracking Measures:} We recorded participants' gaze using a Tobii Pro Fusion eye tracker at a sampling rate of 120 Hz. Pupil diameter, total fixation duration, and saccade count were extracted using Tobii Pro Lab software. These metrics serve as real-time, implicit indicators of attentional and cognitive load~\cite{kosch2023survey, ke2024using, skaramagkas2021review}. As discussed in Section~\ref{sec:background}, pupil diameter reflects cognitive effort driven by the autonomic nervous system, while fixation duration and saccade count are more directly reflective of how attention is distributed across the news interface. Eye-tracking metrics were computed over the full reading block for each article, including both the disclosure statement and the article body. This means that the metrics capture the combined attentional and cognitive effects of reading the disclosure and subsequently processing the article, rather than isolating the disclosure reading alone.

\textbf{Semi-Structured Interview:} We also conducted semi-structured interviews after each session to capture participants' qualitative impressions of the articles and disclosures as well as broader perceptions of AI in journalism and disclosure preferences (details in Section~\ref{sec:procedure}). All interviews were recorded and transcribed.

\subsection{Procedure}
\label{sec:procedure}

The experiment began with informed consent and a description of the news-reading interface. The eye tracker was then calibrated using Tobii Pro Lab's five-point calibration procedure, where participants followed a series of dots on the screen. After calibration, participants accessed a locally hosted web interface where they read the news articles and completed the associated questionnaires. On the first page, they provided demographic information, including age range, gender, and news consumption habits.

As shown in Figure~\ref{fig:loadprotocol}, the study was structured into three sessions (S1, S2, S3), each consisting of four articles (two political, two lifestyle). The articles for each session were assigned using a Latin square rotation, ensuring that across participants, each article appeared in all three disclosure conditions. The order of articles within each session was randomized but alternated between news types. Participants were randomly assigned to Group A or Group B. Each article sequence generated by the Latin square rotation was assigned to one participant from Group A and one from Group B, ensuring that paired participants read identical articles in identical order, with only the disclosure statements differing between them. 

Both groups read articles without any AI disclosure in S1. The no-disclosure condition was always presented first so that participants would read those articles without prior knowledge of AI involvement in the study. In S2 and S3, Group A read articles with one-line disclosures while Group B read articles with detailed disclosures. Participants were instructed to read the disclosure before proceeding to the article. All articles in S2 were either AI-edited or partially AI-generated, with the alternative versions presented in S3, counterbalanced across participants.

After each article, participants completed the NASA-TLX sub-scales. A short semi-structured interview was conducted after each session to capture participants' immediate impressions of the articles and disclosures. After the final session, a longer interview (15--20 minutes) was conducted, which focused on broader perceptions of AI in journalism and disclosure preferences. The eye tracker recorded gaze data throughout the experiment. Each session lasted approximately 10 minutes, and the entire experiment lasted about one hour.

\subsection{Participants}

We recruited 40 participants %\footnote{A power analysis using G*Power indicated that 36 participants would be sufficient for a repeated measures within-between factors F-test to detect a medium effect size (f = 0.25) with $\alpha$ = 0.05 and a power ($1 - \beta$) of 0.95.} 
from a research institute campus, including students, researchers, and non-scientific staff, and were compensated 10 Euros for their time. The study was approved by the institute's ethical committee.

During the post-session interviews, six participants (3 from each group) reported that they did not notice or read the disclosure statement in at least one session. Since these participants' responses could not be attributed to the disclosure condition, they were excluded from all further analyses, resulting in a final sample of 34 participants (21 male, 12 female, 1 non-binary). The majority of participants (n=22, 64.7\%) were aged 25-34, with the remaining participants distributed across the 18-24 (n=2), 35-44 (n=4), 45-54 (n=2), 55-64 (n=2), and 65+ (n=2) age ranges. Most participants (n=19, 55.9\%) reported consuming news multiple times a day, primarily via social media (38.2\%) and online news websites (35.3\%). Additionally, eye-tracking data were missing for four participants in some segments; these participants were excluded from eye-tracking analyses, resulting in 30 participants for eye-tracking measures.

Participants also reported high generative AI literacy (M=26.21, SD=3.19, out of 30), measured using six 5-point Likert scale items from \citeauthor{chan2023students}~\cite{chan2023students}. This relatively homogeneous and high AI literacy is consistent with the campus-based sample and is acknowledged as a limitation.

%Six participants reported not noticing or reading the disclosure in at least one session and were excluded from analyses, resulting in a final sample of 34 participants. The majority of participants were aged 25-34 (N = 22, 64.7\%), with smaller numbers in other age ranges. Twenty-one participants identified as male (61.8\%), 12 as female (35.3\%), and one as non-binary (2.9\%). Most participants (N = 19, 55.9\%) reported consuming news multiple times a day, with smaller proportions reading daily (N=5, 14.7\%) or a few times a week (N=6, 17.6\%). Primary news sources included social media (N=13, 38.2\%) and online news websites (N=12, 35.3\%), with fewer relying on video platforms, news aggregator apps, podcasts/radio, print, or other sources.

\section{Results}

We computed means and standard deviations for NASA-TLX scores and eye-tracking measures (pupil diameter, total fixation duration, saccade count). Statistical analyses using GLMs were conducted for measures where descriptive statistics indicated meaningful variation across conditions; measures with negligible variation (e.g., pupil diameter) were not tested as they would lack practical significance. Given the exploratory nature of the study, p-values were adjusted using the Benjamini-Hochberg procedure~\cite{benjamini1995controlling} to control for false discoveries due to multiple comparisons. Effect sizes (Cohen's d), z-values, and adjusted p-values are reported for pairwise comparisons.

\begin{figure*}
    \centering
    \begin{subfigure}{\textwidth}
        \centering
        \includegraphics[width=0.85\linewidth]{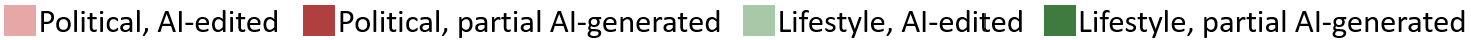}
    \end{subfigure}
    \vspace{1pt}
    \begin{subfigure}{.49\textwidth}
        \centering
        \includegraphics[width=\linewidth]{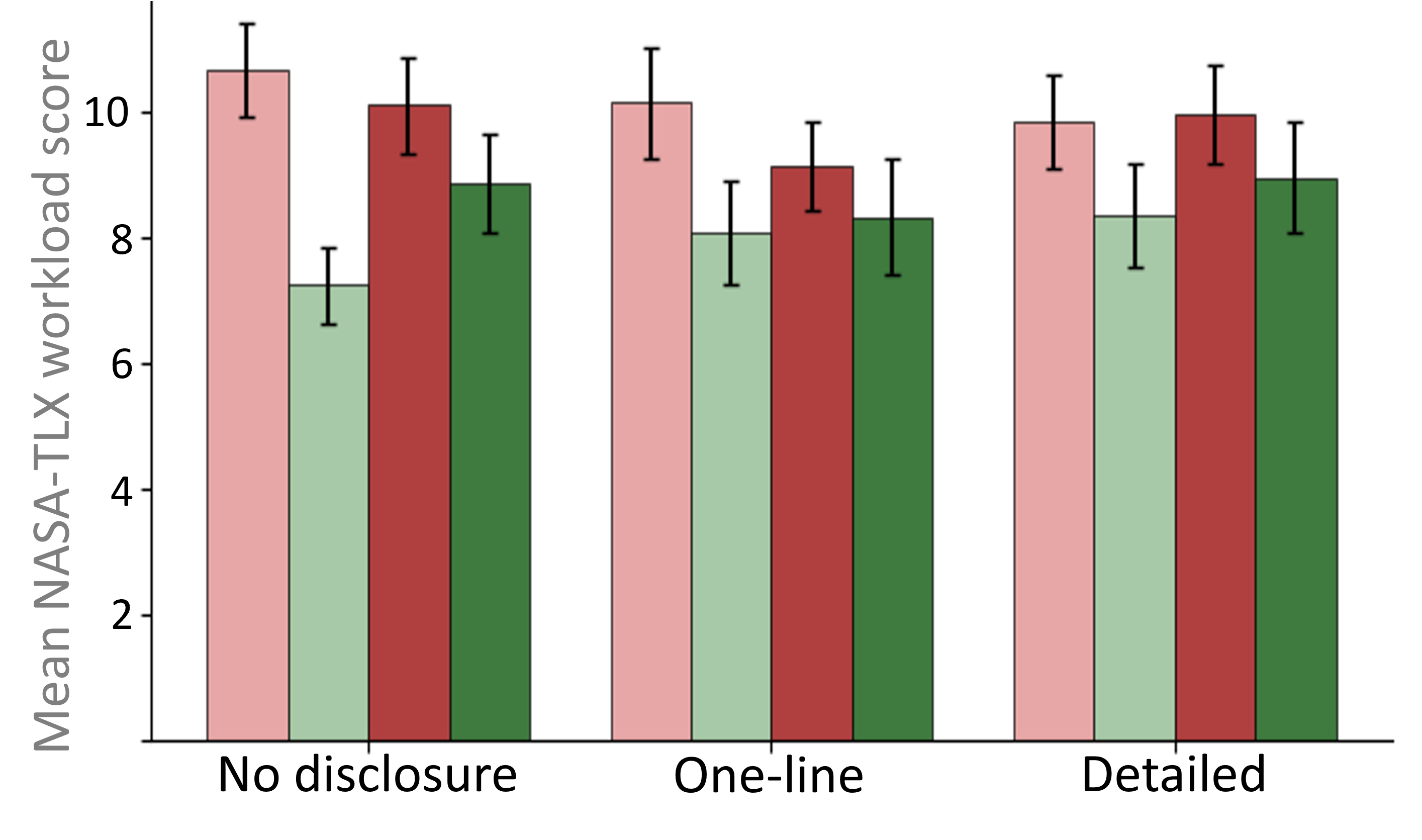}
        %\vspace{-10pt}
        \caption{}
        \label{fig:tlx}
    \end{subfigure}
    \hspace{5pt}
    \begin{subfigure}{.49\textwidth}
        \centering
        \includegraphics[width=\linewidth]{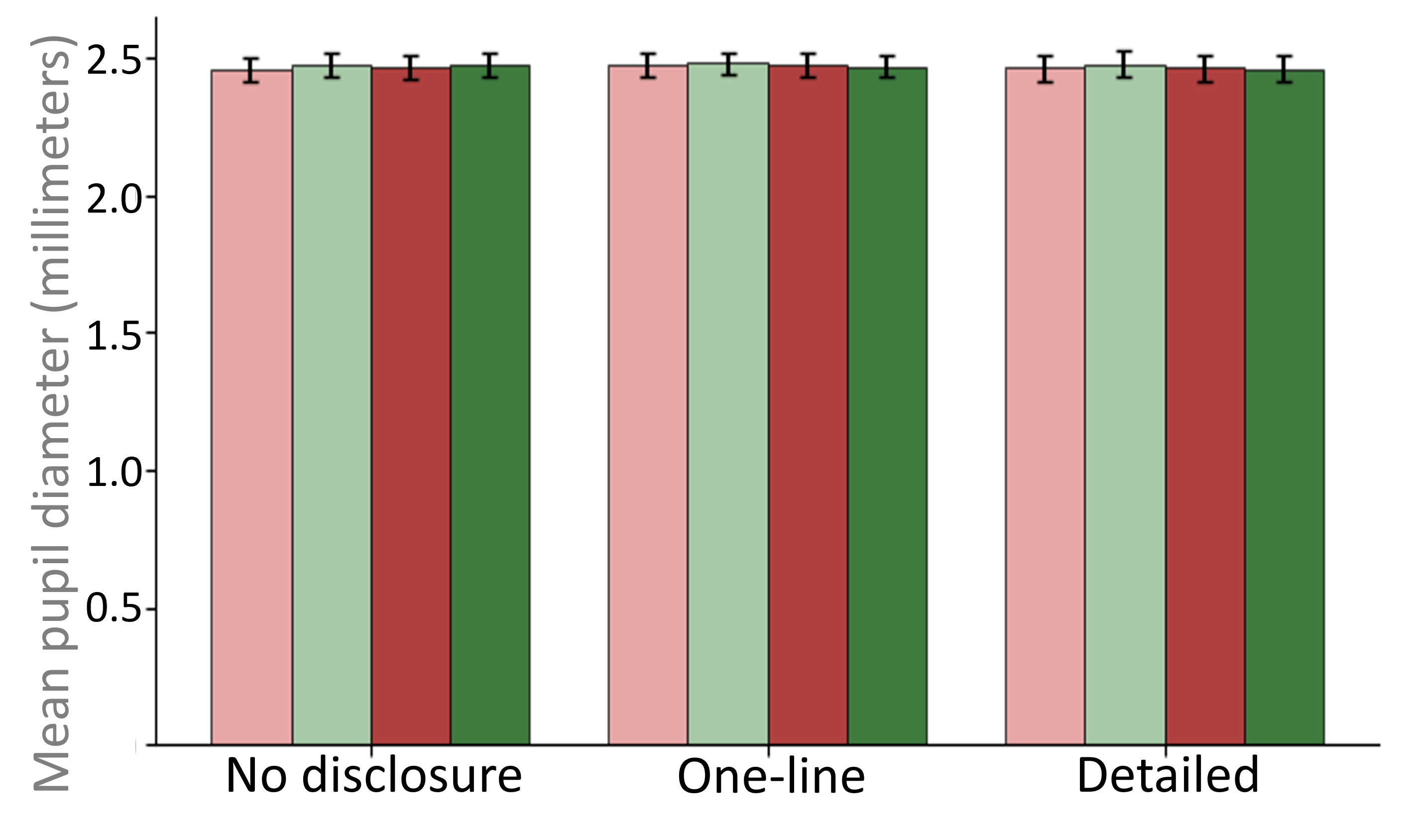}
        %\vspace{-10pt}
            \caption{}
            \label{fig:pupil}
    \end{subfigure}
    \vspace{1pt}
    \begin{subfigure}{.49\textwidth}
        \centering
        \includegraphics[width=\linewidth]{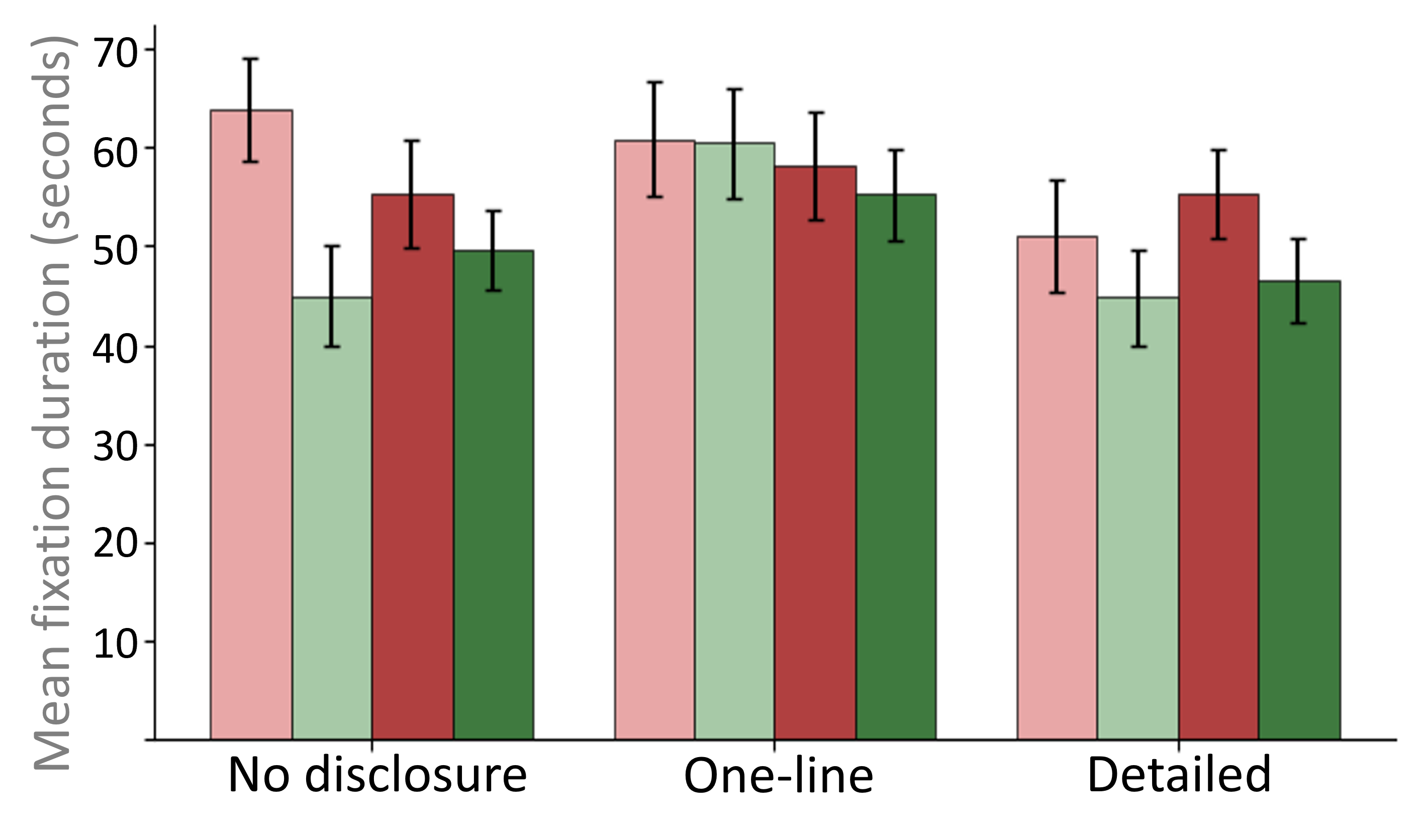}
        %\vspace{-10pt}
        \caption{}
        \label{fig:fixation}
    \end{subfigure}
    \hspace{5pt}
    \begin{subfigure}{.49\textwidth}
        \centering
        \includegraphics[width=\linewidth]{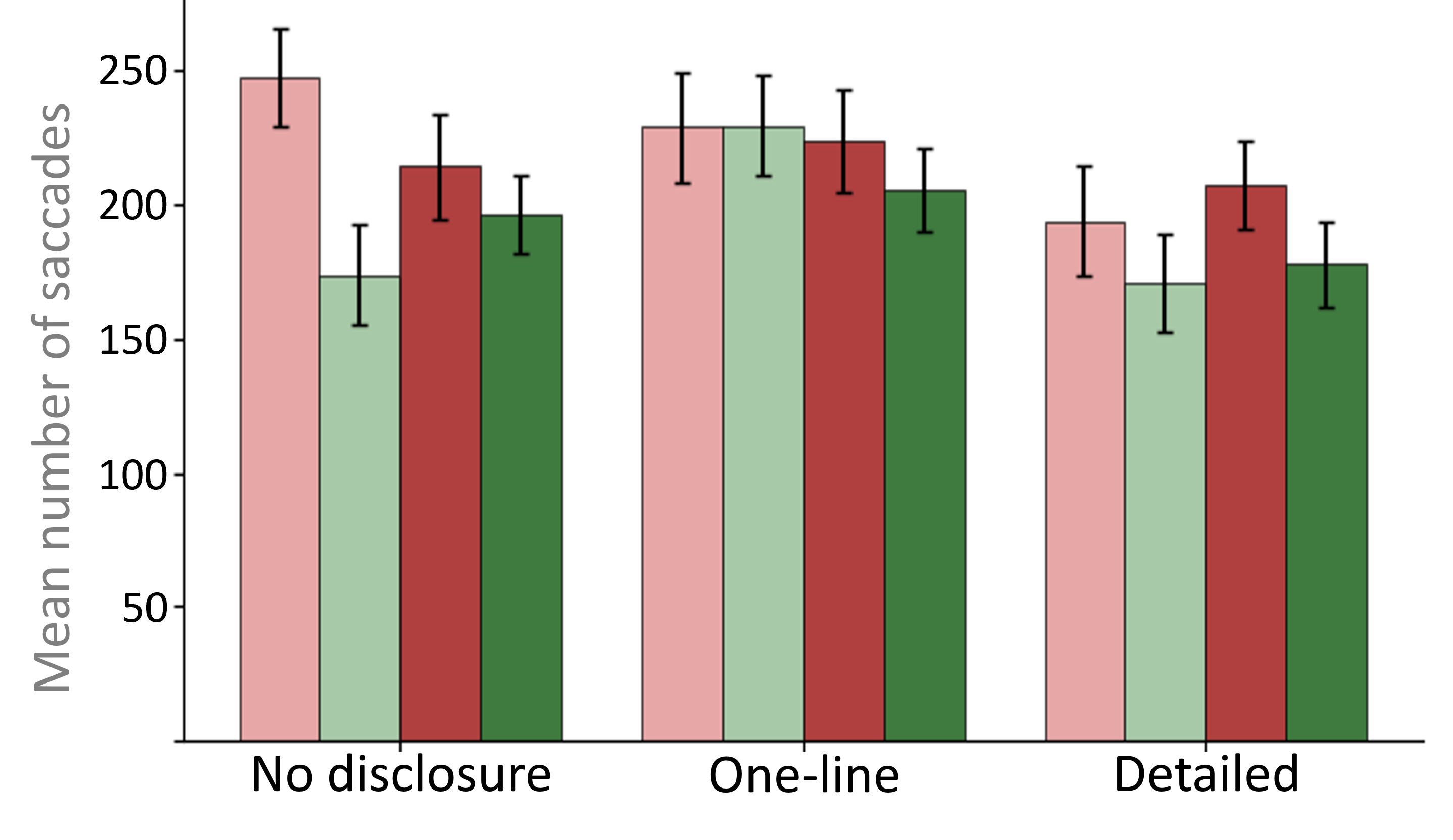}
        %\vspace{-10pt}
            \caption{}
            \label{fig:saccades}
    \end{subfigure}%
    %\vspace{-10pt}
    \caption{Grouped bar visualization of (a) mean NASA-TLX workload scores, (b) mean pupil diameter, (c) mean total fixation duration, and (d) mean saccade count across the three disclosures. Error bars represent the standard error of the mean.}
    \Description[Four grouped bar charts showing (a) mean NASA-TLX scores, (b) mean pupil diameter, (c) average total fixation duration, and (d) average saccade count]{Four grouped bar charts comparing no disclosure, one-line, and detailed disclosure conditions across four measures. NASA-TLX and pupil diameter are similar across all conditions, while fixation duration and saccade count are highest under one-line disclosures and lowest under detailed disclosures. Each condition is broken down by news type (political, lifestyle) and AI role (AI-edited, partial AI-generated).}
\end{figure*}

\subsection{NASA-TLX and Pupil Diameter}

NASA-TLX scores (Figure~\ref{fig:tlx}) were similar across disclosure conditions (no disclosure: 9.21; one-line: 8.91; detailed: 9.26). In all conditions, political articles elicited higher task load than lifestyle articles (e.g., no disclosure: political 10.37, lifestyle 8.04), consistent with expectations for hard versus soft news. Across all disclosure conditions, task load was similar for AI-edited (no disclosure: 8.94, one-line: 9.09, detailed: 9.08) and partially AI-generated (no disclosure: 9.47, one-line: 8.72, detailed: 9.44) articles. This suggests that participants did not perceive a difference in cognitive effort between reading AI-edited and partially AI-generated content, regardless of whether and how AI use was disclosed.

Similarly, pupil diameter (Figure~\ref{fig:pupil}) was consistent across all disclosure conditions (no disclosure: 2.469; one-line: 2.477; detailed: 2.468). Unlike NASA-TLX, pupil diameter did not show notable differences between political and lifestyle articles (e.g., no disclosure: political 2.465, lifestyle 2.473). Disclosing AI involvement also did not affect pupil diameter (AI-edited: no disclosure 2.468, one-line 2.480, detailed 2.471; partially AI-generated: no disclosure 2.470, one-line 2.473, detailed 2.464).

\vspace{3pt}
\noindent$\blacktriangleright$\textbf{Takeaway:} \textit{Both subjective (NASA-TLX) and eye-tracking (pupil diameter) indicators of cognitive load showed no differences across disclosure conditions, suggesting that AI-use disclosures do not impose additional cognitive burden regardless of detail level.}
%\vspace{3pt}

\subsection{Fixation Duration and Saccade Count}
Unlike cognitive load measures, fixation duration and saccade count showed promising differences across disclosure conditions.

Fixation durations (Figure~\ref{fig:fixation}) were highest under one-line disclosures and lowest under detailed disclosures (no: 53.46, one-line: 58.70, detailed: 49.46). One-line disclosures led to significantly higher fixation durations compared to detailed disclosures overall (z = 2.544, d = 0.34, adjusted-p = 0.02). This difference was primarily driven by AI-edited articles (no: 54.40, one-line: 60.66, detailed: 47.97), where the difference between one-line and detailed conditions was significant (z = 2.443, d = 0.43, adjusted-p = 0.02). The effect was less pronounced for partially AI-generated articles (no: 52.52, one-line: 56.74, detailed: 50.95), where fixation durations were more similar across disclosure conditions. One-line disclosures also led to higher fixation durations in lifestyle articles (no: 47.32, one-line: 57.84, detailed: 45.75), with significant differences for one-line vs. no disclosure (z = 2.468, d = 0.41, adjusted-p = 0.02) and one-line vs. detailed (z = 2.120, d = 0.46, adjusted-p = 0.039). Although political articles showed higher fixation durations overall (e.g., no: political 59.60, lifestyle 47.32), the differences between disclosure conditions were less prominent for political articles compared to lifestyle articles. In other words, this difference was primarily driven by news type rather than disclosures.

Saccade counts (Figure~\ref{fig:saccades}) followed an identical pattern (no: 207.63, one-line: 221.62, detailed: 187.33). One-line disclosures led to significantly higher saccade counts compared to detailed disclosures overall (z = 2.655, d = 0.35, adjusted-p = 0.02). In AI-edited articles (no: 210.33, one-line: 228.93, detailed: 182.37), the difference between one-line and detailed was significant (z = 2.518, d = 0.44, adjusted-p = 0.02). As with fixation duration, the effect was more pronounced for AI-edited articles than for partially AI-generated articles (no: 204.93, one-line: 214.30, detailed: 192.28). In lifestyle articles (no: 184.70, one-line: 217.10, detailed: 174.27), the difference was significant for one-line vs. detailed (z = 2.460, d = 0.46, adjusted-p = 0.02) and trending towards significance for one-line vs. no disclosure (z = 1.817, d = 0.35, adjusted-p = 0.069). Political articles again showed higher saccade counts overall (e.g., no disclosure: political 230.57, lifestyle 184.70), indicating greater attentional demands of hard news content rather than disclosure effects.

Given the brevity of the one-line disclosure, the observed differences in total fixation duration and saccade count cannot plausibly be attributed to repeated re-reading of the disclosure text alone and likely reflect altered processing of the article body. Furthermore, despite AI-edited articles being easier to read than partially AI-generated articles (Table~\ref{tab:readability}), one-line disclosures led to significantly higher attentional load specifically for AI-edited articles, indicating that the observation stems from disclosures rather than text complexity. 

\vspace{3pt}
\noindent$\blacktriangleright$\textbf{Takeaway:} \textit{One-line disclosures increased attentional load, as reflected in both fixation duration and saccade count, particularly for AI-edited and lifestyle articles. Detailed disclosures did not impose additional attentional burden.}
%\vspace{3pt}

\subsection{Interview Insights}

The interviews covered participants' broader perceptions of AI in journalism, accountability, and disclosure designs. %An inductive thematic analysis with three coders is reported in *blinded for review*. %a companion paper~\cite{prajod2026full}. 
Interviews were analyzed inductively~\cite{braun2006using} by three coders.
Here, we highlight insights relevant to interpreting the attentional and cognitive load findings.

Several participants noted that one-line disclosures were ambiguous about AI's specific role, leaving room for different interpretations. For instance, one participant remarked: ``[One-line disclosure] \textit{was a bit general, I don't know if my mother would read an article and read AI was used as a final layer, I don't know what she would understand from that. So I think it's a bit specific for our generation that knows a bit more.}'' (P3). Participants also mentioned that one-line disclosures lacked information they considered important, such as assurance of human oversight and contact for reporting errors. In contrast, detailed disclosures were perceived as more transparent, giving readers a clearer picture of which steps AI was involved in. However, participants (N=6) noted that detailed disclosures were too long and likely to be skipped, drawing parallels with cookie banners and terms-of-use agreements.

When asked about their disclosure preferences, a majority of participants (N=22) preferred detailed disclosures. Interestingly, many participants (N=15) suggested ideal disclosure designs with a detail-on-demand feature, where a short disclosure is presented by default with the option to obtain more detail. This included three-fourths of the participants who preferred one-line (N=11) disclosures. Suggested designs included a clickable icon next to a short disclosure that expands into a detailed statement, and special icons or codes that link to a public detailed statement on the outlet's website explaining how AI was involved. The main reasoning was that readers should have agency over the level of disclosure and can choose depending on the topic.

Interview insights also suggest that interest and curiosity played a key role in participants' engagement with articles. Participants reported being curious about unfamiliar lifestyle topics, which may have increased engagement and arousal for those articles.

\section{Discussion}

\subsection{Key Findings}
Our findings show that AI-use disclosures influence readers' attentional load, even when cognitive load remains largely unchanged. Eye-tracking measures revealed differences across disclosure conditions that subjective measures did not capture, highlighting the value of real-time gaze sensing for evaluating disclosure designs.

One-line disclosures imposed the highest attentional burden, as reflected in both fixation duration and saccade count, an effect particularly pronounced for AI-edited articles. Detailed disclosures, despite containing more information, did not differ substantially from the no-disclosure baseline. This observation is in line with the Information-Gap theory~\cite{loewenstein1994psychology}. One-line disclosures likely led to a 'gap' between readers' awareness of AI use and their understanding of its role. This unresolved uncertainty appears to have triggered active information-seeking and visual search. Interview data further support this interpretation, highlighting that one-line disclosures introduce ambiguity by omitting details about the specific steps in which AI was involved. This can cause readers to spend additional attentional effort to infer the extent and implications of AI's role, similar to a verification process~\cite{sun2025understanding}. In contrast, detailed disclosures appear to reduce this uncertainty by explicitly clarifying AI's role and confirming human oversight, thereby mitigating attentional demands despite presenting more text. Notably, although such eye-tracking patterns may be associated with reading complex texts~\cite{de2023reading, ilyas2025reading}, this effect was pronounced for AI-edited articles despite their easier readability compared to partially AI-generated articles (Flesch readability: AI-edited M=40.18, AI-generated M=26.43; see Table~\ref{tab:readability}), suggesting the effect is driven by disclosure ambiguity rather than text complexity.

The attentional load effect was also more pronounced for lifestyle articles than for political articles. Although political articles elicited higher fixation durations and saccade counts overall, this difference was primarily driven by news type rather than disclosures. For lifestyle articles, where baseline attentional load was lower, the introduction of a one-line disclosure produced a more noticeable increase, suggesting that attentional effects of disclosures may be more salient when readers are processing less demanding content.

NASA-TLX scores and pupil diameter showed no significant differences across conditions, suggesting that AI-use disclosures do not impose cognitive burden regardless of the level of detail. NASA-TLX scores showed higher load for political articles than lifestyle, which is in line with the readability scores. However, pupil diameters were similar across news types, which may be partly explained by interest and curiosity. Interview insights suggest that participants were less familiar with some lifestyle topics, which led to unexpected curiosity and interest. Arousal stemming from curiosity/interest may have led to higher pupil dilation~\cite{brod2019lighting} for lifestyle articles, narrowing the expected difference with cognitively demanding political articles.

Higher attentional load from one-line disclosures may have practical consequences beyond the reading experience. Prior work has shown that higher attentional demands during news reading can reduce information retention and increase susceptibility to misinformation~\cite{debue2014does, apuke2024information}. Whether the attentional load observed in our study translates to such downstream effects remains an open question for future work.

The divergence between attentional and cognitive load measures has methodological implications for the HCI community. Subjective measures and physiological indicators of cognitive load (pupil diameter) were not sensitive to disclosure-induced processing differences, whereas attentional measures (fixation duration, saccade count) were. This underscores that attentional and cognitive load are disconnected constructs in this context: disclosures can affect how readers distribute their attention without increasing their cognitive load. For interface designers, this means that relying solely on subjective measures or arousal indicators may miss meaningful differences in how users process transparency information.

\subsection{Implications for Adaptive Disclosure Design}

Our findings have direct implications for the design of adaptive disclosure interfaces. While detailed disclosures perform better on attentional measures, participants noted they would pay less attention to lengthy disclosures over time, drawing parallels with cookie banners~\cite{el2024transparent, prajod2026full}. This tension motivates detail-on-demand disclosure designs,  where a concise one-line disclosure is presented by default with the option to expand for more detail~\cite{springer2020progressive, kusters2026human}. Gaze signals provide empirical grounding for such adaptive interfaces, potentially triggering progressive disclosures when a higher attentional load is detected. Such interfaces could dynamically balance transparency and attentional burden based on readers' real-time gaze patterns and news context.

Connecting to the broader AI disclosure literature, studies have found that more transparency or detailed disclosures tend to reduce trust~\cite{prajod2026full, ngo2025balancing}. Taken together, these findings present a design challenge: detailed disclosures reduce attentional load but risk reducing trust, while one-line disclosures maintain trust but increase attentional load. Detail-on-demand designs, informed by real-time gaze sensing, represent a promising path to navigate this trade-off.

\subsection{Limitations and Future Work}
This study has several limitations. First, the participant sample was campus-based and reported high, relatively homogeneous AI literacy scores (M=26.21, SD=3.19, out of 30), which limits generalizability to broader populations with varying levels of AI familiarity. Readers with lower AI literacy may respond differently to AI-use disclosures, and future work should investigate this with more diverse samples.

Second, the lab-based setting differs from real-world news consumption, which typically involves skimming, multitasking, and reading across multiple devices, reducing ecological validity. The controlled environment may have also encouraged more careful reading than would occur naturally.

Third, the no-disclosure condition was always presented first to avoid priming participants about AI involvement. However, the disclosure detail was manipulated as a between-subjects factor. Consequently, any effects of fatigue or interface familiarity would be distributed equally across the one-line and detailed disclosure groups. Since our primary findings emerge from the comparison between these independent groups, the ordering of the baseline does not compromise the validity of the disclosure-detail effects.

Fourth, eye-tracking metrics were computed over the full reading block, including both the disclosure text and article body. While this captures the holistic user experience, future work could employ Area of Interest (AOI) analysis to isolate the specific contributions of the disclosure region. However, we note that detailed disclosures resulted in lower overall attentional load, despite containing more text than one-line disclosures. Since our findings cannot be explained by disclosure lengths, they are not a result of simple disclosure reading time.

Future work should investigate detail-on-demand disclosure designs and explore how gaze-based adaptive disclosures can be implemented and evaluated in news interfaces. Building and testing a prototype that uses real-time gaze signals to trigger progressive disclosures would be a natural next step. Additionally, longitudinal studies examining how readers' attentional responses to disclosures change with repeated exposure would provide insights into habituation effects relevant to real-world deployment.

\section{Conclusion}
We investigated how AI-use disclosure detail level and AI's role in content production affect readers' attentional and cognitive load through a $3\times2\times2$ mixed-factorial study, combining NASA-TLX and eye-tracking measures. One-line disclosures increased attentional load, as reflected in higher fixation durations and saccade counts, particularly for AI-edited and lifestyle articles. Detailed disclosures did not impose additional burden despite containing more information. NASA-TLX scores and pupil diameter showed no differences across conditions, indicating that cognitive load remains unchanged regardless of disclosure detail. Interview data, together with Information-Gap Theory, suggest that the ambiguity of one-line disclosures is a plausible reason for higher attentional load, while detailed disclosures reduce uncertainty by clarifying AI's role. Furthermore, participants' ideal disclosure involved detail-on-demand designs, reinforcing the need for adaptive approaches that balance transparency with attentional burden. Our findings demonstrate the value of eye-tracking for evaluating disclosure designs and motivate gaze-informed adaptive disclosure interfaces that dynamically adjust detail level based on readers' attentional patterns and news context.

\section*{Safe and Responsible Innovation Statement}
This research promotes responsible innovation in AI transparency by investigating non-invasive eye-tracking measures in a controlled lab setting. No deception was used; participants were fully informed about the study before providing their written informed consent, and were free to withdraw at any time. The study was approved by an ethical committee, and all data were stored anonymously in accordance with GDPR guidelines. Potential biases due to the campus-based sample are acknowledged, and future work aims to address this through more diverse populations. In real-world applications of gaze-based adaptive disclosure interfaces, careful attention would be necessary to ensure inclusivity, prevent dark patterns, and avoid overreliance on automated disclosures.

\section*{Acknowledgment}
Except for creating the news article stimuli, generative AI (Claude Opus 4.6) was used only for rephrasing parts of this manuscript. %This publication is part of the AI, Media \& Democracy Lab (Dutch Research Council project number: NWA.1332.20.009).

\bibliographystyle{ACM-Reference-Format}
\bibliography{chi2026_main}

@article{becker2025policies,
  title={Policies in parallel? A comparative study of journalistic AI policies in 52 global news organisations},
  author={Becker, Kim Bj{\"o}rn and Simon, Felix M and Crum, Christopher},
  journal={Digital Journalism},
  volume={13},
  number={9},
  pages={1578--1598},
  year={2025},
  publisher={Taylor \& Francis}
}

@inproceedings{zier2024labeling,
  title={Labeling AI-Generated News Content: Matching Journalist Intentions with Audience Expectations},
  author={Zier, Jessica and Diakopoulos, Nicholas},
  booktitle={Proceedings of the Computafion and Journalism Symposium 2024},
  year={2024}
}

@inproceedings{morosoli2025transparency,
  title={The Transparency Dilemma: An Experiment on How AI Disclosures Affect Credibility Perceptions and Engagement Across Topics},
  author={Morosoli, Sophie and van der Goot, Emma and Resendez, Valeria and de Vreese, Claes and Helberger, Natali},
  booktitle={Proceedings of the AAAI/ACM Conference on AI, Ethics, and Society},
  volume={8},
  number={2},
  pages={1748--1757},
  year={2025}
}

@article{nanz2025ai,
  title={AI in the Newsroom: Does the Public Trust Automated Journalism and Will They Pay for It?},
  author={Nanz, Andreas and Binder, Alice and Matthes, J{\"o}rg},
  journal={Journalism Studies},
  pages={1--20},
  year={2025},
  publisher={Taylor \& Francis}
}

@article{altay2024people,
  title={People are skeptical of headlines labeled as AI-generated, even if true or human-made, because they assume full AI automation},
  author={Altay, Sacha and Gilardi, Fabrizio},
  journal={PNAS nexus},
  volume={3},
  number={10},
  pages={pgae403},
  year={2024},
  publisher={Oxford University Press US}
}

@article{toff2025or,
  title={“Or they could just not use it?”: The dilemma of AI disclosure for audience trust in news},
  author={Toff, Benjamin and Simon, Felix M},
  journal={The International Journal of Press/Politics},
  volume={30},
  number={4},
  pages={881--903},
  year={2025},
  publisher={SAGE Publications Sage CA: Los Angeles, CA}
}

@inproceedings{longoni2022news,
  title={News from generative artificial intelligence is believed less},
  author={Longoni, Chiara and Fradkin, Andrey and Cian, Luca and Pennycook, Gordon},
  booktitle={Proceedings of the 2022 ACM Conference on Fairness, Accountability, and Transparency},
  pages={97--106},
  year={2022}
}

@article{prajod2026full,
  title={Full Disclosure, Less Trust? How the Level of Detail about AI Use in News Writing Affects Readers' Trust},
  author={Prajod, Pooja and Cools, Hannes and R{\"o}ggla, Thomas and Venkatraj, Karthikeya Puttur and Kusters, Amber and ElKattan, Alia and Cesar, Pablo and Ali, Abdallah El},
  journal={arXiv preprint arXiv:2601.09620},
  year={2026}
}

@article{liu2023nudging,
  title={The nudging effect of AIGC labeling on users’ perceptions of automated news: evidence from EEG},
  author={Liu, Yuhan and Wang, Shuining and Yu, Guoming},
  journal={Frontiers in Psychology},
  volume={14},
  pages={1277829},
  year={2023},
  publisher={Frontiers Media SA}
}

@inproceedings{shi2023true,
  title={True or false? Cognitive load when reading COVID-19 news headlines: an eye-tracking study},
  author={Shi, Li and Bhattacharya, Nilavra and Das, Anubrata and Gwizdka, Jacek},
  booktitle={Proceedings of the 2023 Conference on Human Information Interaction and Retrieval},
  pages={107--116},
  year={2023}
}

@article{zhang2024you,
  title={You Can't Beat the Affects: How Emotional News Written by AI Affect the Psychophysiological Responses of Young Readers},
  author={Zhang, Yan and Xu, Yawen and Li, Li and Wen, Tianyi and Ding, Jie},
  journal={Available at SSRN 4950484},
  year={2024}
}

@inproceedings{ilyas2025reading,
  title={Reading the Readers Mind through Eye Tracking: Can AI Generated Texts Match Human Authors?},
  author={Ilyas, Chaudhary Muhammad Aqdus and Noor, Sifat-E and Tashk, Ashkan and Cooreman, Bart and Beier, Sofie and B{\ae}kgaard, Per},
  booktitle={Proceedings of the 2025 Symposium on Eye Tracking Research and Applications},
  pages={1--7},
  year={2025}
}

@incollection{hart1988development,
  title={Development of NASA-TLX (Task Load Index): Results of empirical and theoretical research},
  author={Hart, Sandra G and Staveland, Lowell E},
  booktitle={Advances in psychology},
  volume={52},
  pages={139--183},
  year={1988},
  publisher={Elsevier}
}

@inproceedings{el2024transparent,
  title={Transparent AI disclosure obligations: Who, what, when, where, why, how},
  author={El Ali, Abdallah and Venkatraj, Karthikeya Puttur and Morosoli, Sophie and Naudts, Laurens and Helberger, Natali and Cesar, Pablo},
  booktitle={Extended Abstracts of the CHI Conference on Human Factors in Computing Systems},
  pages={1--11},
  year={2024}
}

@article{springer2020progressive,
  title={Progressive disclosure: When, why, and how do users want algorithmic transparency information?},
  author={Springer, Aaron and Whittaker, Steve},
  journal={ACM Transactions on Interactive Intelligent Systems (TiiS)},
  volume={10},
  number={4},
  pages={1--32},
  year={2020},
  publisher={ACM New York, NY, USA}
}

@article{ngo2025balancing,
  title={Balancing AI transparency: Trust, Certainty, and Adoption},
  author={Ngo, Vu Minh},
  journal={Information Development},
  pages={02666669251346124},
  year={2025},
  publisher={SAGE Publications Sage UK: London, England}
}

@article{hossain2025effects,
  title={Effects of Algorithmic Transparency on User Experience and Physiological Responses in Affect-Aware Task Adaptation},
  author={Hossain, Mohammad Sohorab and Clapp, Joshua D and Novak, Vesna D},
  journal={IEEE Transactions on Affective Computing},
  year={2025},
  publisher={IEEE}
}

@inproceedings{gamage2025labeling,
  title={Labeling Synthetic Content: User Perceptions of Label Designs for AI-Generated Content on Social Media},
  author={Gamage, Dilrukshi and Sewwandi, Dilki and Zhang, Min and Bandara, Arosha K},
  booktitle={Proceedings of the 2025 CHI Conference on Human Factors in Computing Systems},
  pages={1--29},
  year={2025}
}

@inproceedings{kusters2026human,
  title={More Human or More AI? Visualizing Human-AI Collaboration Disclosures in Journalistic News Production},
  author={Kusters, Amber and Prajod, Pooja and Cesar, Pablo and El Ali, Abdallah},
  booktitle={Proceedings of the 2026 CHI Conference on Human Factors in Computing Systems},
  pages={1--22},
  year={2026}
}

@misc{longdisclosure,
  title={The Disclosure Dilemma: How {AI} Attribution Affects Reactions to Public Health Messages},
  author={Long, Jacob A. and Oyewole, Tabitha and Goli, Maryam and Keisler, Jacqueline M. and Alyaqout, Saud and Rodgers, Michael D. and {N'Diaye}, Arielle},
  howpublished={Poster presented at the 108th Annual Conference of the Association for Education in Journalism and Mass Communication (AEJMC)},
  year={2025}
}

@article{kosch2023survey,
  title={A survey on measuring cognitive workload in human-computer interaction},
  author={Kosch, Thomas and Karolus, Jakob and Zagermann, Johannes and Reiterer, Harald and Schmidt, Albrecht and Wo{\'z}niak, Pawe{\l} W},
  journal={ACM Computing Surveys},
  volume={55},
  number={13s},
  pages={1--39},
  year={2023},
  publisher={ACM New York, NY}
}

@article{ke2024using,
  title={Using eye-tracking in education: review of empirical research and technology},
  author={Ke, Fengfeng and Liu, Ruohan and Sokolikj, Zlatko and Dahlstrom-Hakki, Ibrahim and Israel, Maya},
  journal={Educational technology research and development},
  volume={72},
  number={3},
  pages={1383--1418},
  year={2024},
  publisher={Springer}
}

@article{skaramagkas2021review,
  title={Review of eye tracking metrics involved in emotional and cognitive processes},
  author={Skaramagkas, Vasileios and Giannakakis, Giorgos and Ktistakis, Emmanouil and Manousos, Dimitris and Karatzanis, Ioannis and Tachos, Nikolaos S and Tripoliti, Evanthia and Marias, Kostas and Fotiadis, Dimitrios I and Tsiknakis, Manolis},
  journal={IEEE reviews in biomedical engineering},
  volume={16},
  pages={260--277},
  year={2021},
  publisher={IEEE}
}

@article{benjamini1995controlling,
  title={Controlling the false discovery rate: a practical and powerful approach to multiple testing},
  author={Benjamini, Yoav and Hochberg, Yosef},
  journal={Journal of the Royal statistical society: series B (Methodological)},
  volume={57},
  number={1},
  pages={289--300},
  year={1995},
  publisher={Wiley Online Library}
}

@inproceedings{chiossi2024physiochi,
  title={PhysioCHI: Towards Best Practices for Integrating Physiological Signals in HCI},
  author={Chiossi, Francesco and Stepanova, Ekaterina R and Tag, Benjamin and Perusquia-Hernandez, Monica and Kitson, Alexandra and Dey, Arindam and Mayer, Sven and El Ali, Abdallah},
  booktitle={Extended Abstracts of the CHI Conference on Human Factors in Computing Systems},
  pages={1--7},
  year={2024}
}

@article{apuke2024information,
  title={Information overload and misinformation sharing behaviour of social media users: Testing the moderating role of cognitive ability},
  author={Apuke, Oberiri Destiny and Omar, Bahiyah and Tunca, Elif Asude and Gever, Celestine Verlumun},
  journal={Journal of Information Science},
  volume={50},
  number={6},
  pages={1371--1381},
  year={2024},
  publisher={SAGE Publications Sage UK: London, England}
}

@article{debue2014does,
  title={What does germane load mean? An empirical contribution to the cognitive load theory},
  author={Debue, Nicolas and Van De Leemput, C{\'e}cile},
  journal={Frontiers in psychology},
  volume={5},
  pages={1099},
  year={2014},
  publisher={Frontiers Media SA}
}

@article{de2023reading,
  title={Reading numbers is harder than reading words: An eye-tracking study},
  author={de Chambrier, Anne-Fran{\c{c}}oise and Pedrotti, Marco and Ruggeri, Paolo and Dewi, Jasinta and Atzemian, Myrto and Thevenot, Catherine and Martinet, Catherine and Terrier, Philippe},
  journal={Acta Psychologica},
  volume={237},
  pages={103942},
  year={2023},
  publisher={Elsevier}
}

@article{sun2025understanding,
  title={Understanding trust toward human versus AI-generated health information through behavioral and physiological sensing},
  author={Sun, Xin and Ma, Rongjun and Wei, Shu and Cesar, Pablo and Bosch, Jos A and El Ali, Abdallah},
  journal={International Journal of Human-Computer Studies},
  pages={103714},
  year={2025},
  publisher={Elsevier}
}

@article{gilardi2024willingness,
  title={Willingness to Read AI-Generated News Is Not Driven by Their Perceived Quality},
  author={Gilardi, Fabrizio and Di Lorenzo, Sabrina and Ezzaini, Juri and Santa, Beryl and Streiff, Benjamin and Zurfluh, Eric and Hoes, Emma},
  journal={arXiv preprint arXiv:2409.03500},
  year={2024}
}

@article{leuppert2025ai,
  title={AI-reporters as the future of journalism? Investigating recipients’ credibility evaluations of AI-authored news},
  author={Leuppert, Robin and Weinmann, Carina and Eiden, Jan},
  journal={Journalism},
  pages={14648849251382484},
  year={2025},
  publisher={SAGE Publications Sage UK: London, England}
}

@inproceedings{prajod2023gaze,
  title={Gaze-based attention recognition for human-robot collaboration},
  author={Prajod, Pooja and Lavit Nicora, Matteo and Malosio, Matteo and Andr{\'e}, Elisabeth},
  booktitle={Proceedings of the 16th International Conference on PErvasive Technologies Related to Assistive Environments},
  pages={140--147},
  year={2023}
}

@article{heimerl2024fordigitstress,
  title={The ForDigitStress dataset: A multi-modal dataset for automatic stress recognition},
  author={Heimerl, Alexander and Prajod, Pooja and Mertes, Silvan and Baur, Tobias and Kraus, Matthias and Liu, Ailin and Risack, Helen and Rohleder, Nicolas and Andr{\'e}, Elisabeth and Becker, Linda},
  journal={IEEE transactions on affective computing},
  volume={16},
  number={2},
  pages={1219--1234},
  year={2024},
  publisher={IEEE}
}

@article{liu2026sensing,
  title={Sensing What Surveys Miss: Understanding and Personalizing Proactive LLM Support by User Modeling},
  author={Liu, Ailin and Karoui, Yesmine and Draxler, Fiona and Kreuter, Frauke and Chiossi, Francesco},
  journal={arXiv preprint arXiv:2602.00880},
  year={2026}
}

@online{BBC,
  author = {BBC},
  title = {How we're designing user-centred AI labels at the BBC
},
  url = {https://www.bbc.co.uk/mediacentre/articles/user-centred-ai-labels},
  urldate = {2026-03-24},
  year = {2025}
}

@online{NordicAI,
  author = {Nordic AI Journalism in collaboration with Utgivarna},
  title = {AI transparency in journalism},
  url = {https://www.nordicaijournalism.com/ai-transparency},
  urldate = {2026-03-24},
  year = {2024}
}

@article{flesch1948new,
  title={A new readability yardstick.},
  author={Flesch, Rudolph},
  journal={Journal of applied psychology},
  volume={32},
  number={3},
  pages={221},
  year={1948},
  publisher={American Psychological Association}
}

@techreport{senter1967automated,
  title={Automated readability index},
  author={Senter, Richard J and Smith, Edgar A},
  year={1967}
}

@article{brod2019lighting,
  title={Lighting the wick in the candle of learning: generating a prediction stimulates curiosity},
  author={Brod, Garvin and Breitwieser, Jasmin},
  journal={NPJ science of learning},
  volume={4},
  number={1},
  pages={17},
  year={2019},
  publisher={Nature Publishing Group UK London}
}

@online{wu2025characterizing,
  title={Characterizing Physiological and Behavioral Responses to Human- and AI-Generated Real and Fake News},
  author={Wu, Lian A. and Prajod, Pooja and El Ali, Abdallah and Cesar, Pablo},
  note={News Futures Workshop at the 2025 CHI Conference on Human Factors in Computing Systems},
  url = {https://sites.google.com/view/newsfutures/accepted-submissions},
  urldate = {2026-04-15},
  year={2025}
}

@inproceedings{chen2025examining,
  title={Examining the Impact of Label Detail and Content Stakes on User Perceptions of AI-Generated Images on Social Media},
  author={Chen, Jingruo and Wang, Tung-Yen and Williams, Marie and Jordan, Natalia Andrea and Shao, Mingyi and Zhang, Linda and Fussell, Susan R},
  booktitle={Companion Publication of the 2025 Conference on Computer-Supported Cooperative Work and Social Computing},
  pages={270--275},
  year={2025}
}

@article{plopski2022eye,
  title={The eye in extended reality: A survey on gaze interaction and eye tracking in head-worn extended reality},
  author={Plopski, Alexander and Hirzle, Teresa and Norouzi, Nahal and Qian, Long and Bruder, Gerd and Langlotz, Tobias},
  journal={ACM Computing Surveys (CSUR)},
  volume={55},
  number={3},
  pages={1--39},
  year={2022},
  publisher={ACM New York, NY}
}

@article{menges2019improving,
  title={Improving user experience of eye tracking-based interaction: Introspecting and adapting interfaces},
  author={Menges, Raphael and Kumar, Chandan and Staab, Steffen},
  journal={ACM Transactions on Computer-Human Interaction (TOCHI)},
  volume={26},
  number={6},
  pages={1--46},
  year={2019},
  publisher={ACM New York, NY, USA}
}

@inproceedings{hansen2020factuality,
  title={Factuality checking in news headlines with eye tracking},
  author={Hansen, Christian and Hansen, Casper and Simonsen, Jakob Grue and Larsen, Birger and Alstrup, Stephen and Lioma, Christina},
  booktitle={Proceedings of the 43rd international ACM SIGIR conference on research and development in information retrieval},
  pages={2013--2016},
  year={2020}
}

@article{sumer2021fakenewsperception,
  title={FakeNewsPerception: An eye movement dataset on the perceived believability of news stories},
  author={S{\"u}mer, {\"O}mer and Bozkir, Efe and K{\"u}bler, Thomas and Gr{\"u}ner, Sven and Utz, Sonja and Kasneci, Enkelejda},
  journal={Data in brief},
  volume={35},
  pages={106909},
  year={2021},
  publisher={Elsevier}
}

@article{helberger2023european,
  title={The European AI act and how it matters for research into AI in media and journalism},
  author={Helberger, Natali and Diakopoulos, Nicholas},
  journal={Digital Journalism},
  volume={11},
  number={9},
  pages={1751--1760},
  year={2023},
  publisher={Taylor \& Francis}
}

@article{chan2023students,
  title={Students’ voices on generative AI: Perceptions, benefits, and challenges in higher education},
  author={Chan, Cecilia Ka Yuk and Hu, Wenjie},
  journal={International journal of educational technology in higher education},
  volume={20},
  number={1},
  pages={43},
  year={2023},
  publisher={Springer}
}

@article{loewenstein1994psychology,
  title={The psychology of curiosity: A review and reinterpretation.},
  author={Loewenstein, George},
  journal={Psychological bulletin},
  volume={116},
  number={1},
  pages={75},
  year={1994},
  publisher={American Psychological Association}
}

@article{braun2006using,
  title={Using thematic analysis in psychology},
  author={Braun, Virginia and Clarke, Victoria},
  journal={Qualitative research in psychology},
  volume={3},
  number={2},
  pages={77--101},
  year={2006},
  publisher={Taylor \& Francis}
}

@article{sun2026eyes,
  title={Eyes Can't Always Tell: Fusing Eye Tracking and User Priors for User Modeling under AI Advice Conditions},
  author={Sun, Xin and Wei, Shu and Pan, Ting and Wang, Yajing and Bosch, Jos A and Echizen, Isao and Ali, Abdallah El and Sugawara, Saku},
  journal={arXiv preprint arXiv:2604.01741},
  year={2026}
}

@article{valenzuela2026effects,
  title={The Effects of Generative AI in News on Media Credibility and Selectivity: Evidence from a Conjoint Experiment in Chile},
  author={Valenzuela, Sebasti{\'a}n and Bachmann, Ingrid and Borah, Porismita and Sol{\'\i}s Vald{\'e}s, Natalia},
  journal={Digital Journalism},
  pages={1--22},
  year={2026},
  publisher={Taylor \& Francis}
}

@inproceedings{barz2024humaneyeze,
  title={HumanEYEze 2024: Workshop on Eye Tracking for Multimodal Human-Centric Computing},
  author={Barz, Michael and Bednarik, Roman and Bulling, Andreas and Conati, Cristina and Sonntag, Daniel},
  booktitle={Proceedings of the 26th International Conference on Multimodal Interaction},
  pages={696--697},
  year={2024}
}

@inproceedings{alhargan2017multimodal,
  title={Multimodal affect recognition in an interactive gaming environment using eye tracking and speech signals},
  author={Alhargan, Ashwaq and Cooke, Neil and Binjammaz, Tareq},
  booktitle={Proceedings of the 19th ACM international conference on multimodal interaction},
  pages={479--486},
  year={2017}
}

@inproceedings{gupta2020eyes,
  title={The eyes know it: Fakeet-an eye-tracking database to understand deepfake perception},
  author={Gupta, Parul and Chugh, Komal and Dhall, Abhinav and Subramanian, Ramanathan},
  booktitle={Proceedings of the 2020 international conference on multimodal interaction},
  pages={519--527},
  year={2020}
}

@inproceedings{murakami2025inferring,
  title={Inferring User State from Gaze Dynamics in a VR Throwing Task: Toward Adaptive and User-Centered Rehabilitation Support},
  author={Murakami, Haruka and Inamura, Tetusnari},
  booktitle={Companion Proceedings of the 27th International Conference on Multimodal Interaction},
  pages={235--239},
  year={2025}
}

@inproceedings{barral2020eye,
  title={Eye-tracking to predict user cognitive abilities and performance for user-adaptive narrative visualizations},
  author={Barral, Oswald and Lall{\'e}, S{\'e}bastien and Guz, Grigorii and Iranpour, Alireza and Conati, Cristina},
  booktitle={Proceedings of the 2020 international conference on multimodal interaction},
  pages={163--173},
  year={2020}
}

@article{lavit2024gaze,
  title={Gaze detection as a social cue to initiate natural human-robot collaboration in an assembly task},
  author={Lavit Nicora, Matteo and Prajod, Pooja and Mondellini, Marta and Tauro, Giovanni and Vertechy, Rocco and Andr{\'e}, Elisabeth and Malosio, Matteo},
  journal={Frontiers in Robotics and AI},
  volume={11},
  pages={1394379},
  year={2024},
  publisher={Frontiers Media SA}
}

@article{prajod2024flow,
  title={Flow in human-robot collaboration—multimodal analysis and perceived challenge detection in industrial scenarios},
  author={Prajod, Pooja and Lavit Nicora, Matteo and Mondellini, Marta and Falerni, Matteo Meregalli and Vertechy, Rocco and Malosio, Matteo and Andr{\'e}, Elisabeth},
  journal={Frontiers in Robotics and AI},
  volume={11},
  pages={1393795},
  year={2024},
  publisher={Frontiers Media SA}
}

\end{document}